\documentclass[12pt]{article}

\usepackage{epsf}
\usepackage{citesort}
\usepackage{epsfig}

\newtheorem{theorem}{Theorem}
\newtheorem{lemma}[theorem]{Lemma}

\newenvironment{proof}{\noindent\par{\bf Proof: }}{\nopagebreak\rule{1
ex}{0.8 em}\medskip}

\newcommand{\KaoThanks}[0]{\thanks{Department of Computer Science, Yale
University, New Haven, CT 06520, USA.
E-mail: {\rm kao-ming-yang@cs.yale.edu}.
Research supported in part by NSF Grant CCR-9531028.}}
\newcommand{\HsuThanks}[0]{\thanks{
Institute of Information Science, Academia Sinica, Nankang 11529,
Taipei, Taiwan, ROC. E-mail: {\rm tshsu@iis.sinica.edu.tw}.
Research supported in part by NSC Grants 85-2213-E-001-003,
86-2213-E-001-012, and 87-2213-E-001-022.}}

\begin{document} 

\title{Optimal Augmentation for Bipartite Componentwise
Biconnectivity in Linear Time}

\author{Tsan-sheng Hsu\HsuThanks\ and Ming-Yang Kao\KaoThanks}

% started April 10, 1996
% final polishing done Aug 24, 1998

\maketitle

\newcommand{\case}[0]{{\it Case}}
\newcommand{\hht}[1]{\Delta(#1)}
\newcommand{\romanzero}[0]{S1}
\newcommand{\romanone}[0]{S2}
\newcommand{\romantwo}[0]{S3}
\newcommand{\romanthree}[0]{S4}
\newcommand{\romanfour}[0]{S5}

\newcommand{\BB}[1]{{\cal B}(#1)}
\newcommand{\CC}[0]{{\cal C}}
\newcommand{\DD}[2]{{\cal D}(#2,#1)}
\newcommand{\bedge}[2]{{<\!#1,#2\!>}}
\newcommand{\MM}[1]{{\cal M}(#1)}
\newcommand{\RR}[1]{{\cal R}(#1)}
\newcommand{\thetree}[1]{\Psi(#1)}
\newcommand{\bound}[0]{\eta}
\newcommand{\thenumber}[1]{\max_{u \in
G}\{\DD{#1}{u}+\CC(#1)-2,\MM{\Lambda(#1)}+\RR{\Lambda(#1)}\}}

%\pagestyle{myheadings}
%\markboth{\sc hsu and kao}{\sc bipartite biconnectivity augmentation}

\begin{abstract}
A graph is {\it componentwise biconnected\/} if every connected
component either is an isolated vertex or is biconnected.  We present
a linear-time algorithm for the problem of adding the smallest number
of edges to make a bipartite graph componentwise biconnected while
preserving its bipartiteness.  This algorithm has immediate
applications for protecting sensitive information in statistical
tables.
\end{abstract}

%\begin{keywords}
%biconnectivity, data security, bipartite graph augmentation
%\end{keywords}

%\begin{AMSMOS}
%68Q20, 68R10, 94C15, 05C40, 05C90
%\end{AMSMOS}

\section{Introduction}
There is a long history of applications for the problem of adding
edges to a graph in order to satisfy connectivity specifications
(see~\cite{fa:network,hts:phd,kg:drawing} for recent examples).
Correspondingly, the problem has been extensively studied for making
general graphs $k$-edge connected or $k$-vertex connected for various
values of $k$
\cite{ET76b,hts:4connect,hts:aug04,hts:biconnect,RG77,wt:vaug} as
well as for making vertex subsets suitably connected
\cite{fa:augmenting,hkao.selaug.conf,ts:augmentation,wt:robust,wt:specified,wt:s3eaug}.

In this paper, we focus on augmenting bipartite graphs.  A graph is
{\it componentwise biconnected\/} if every connected component either
is biconnected or is an isolated vertex.  This paper presents a
linear-time algorithm for the problem of inserting the smallest number
of edges into a given bipartite graph to make it componentwise
biconnected while maintaining its bipartiteness.  This problem and
related bipartite augmentation problems arise naturally from research
on statistical data security
\cite{anr:database,co82,cox80,dde:statistical,kjp:suppression}.  
To protect sensitive information
in a cross tabulated table, it is a common practice to suppress some
of the cells in the table.  A basic issue concerning the effectiveness
of this practice is how a table maker can suppress a small number of
cells in addition to the sensitive ones so that the resulting table
does not leak significant information. This protection problem can be
reduced to augmentation problems for bipartite
graphs~\cite{gusfield88,hkao.marginal.scp,Kao95.augcomp,Kao95.protection,kao95.analytic,mfm:censoring,mfm:suppressing,MMR91}.  
In particular, a linear-time algorithm for our augmentation
problem immediately yields a linear-time algorithm for suppressing the
smallest number of additional cells so that no nontrivial information
about any individual row or column is revealed to an
adversary~\cite{Kao95.protection}.

Section \ref{sec_def} formally states our augmentation problem and
discusses main results.
Section~\ref{sec_special_case} proves an optimal bound on the smallest
number of additional edges needed for the problem.
Section~\ref{sec_linear_time} gives a linear-time algorithm to solve
the augmentation problem.
 
\section{Problem formulation, main results, and basic concepts}\label{sec_def}

In this paper, all graphs are undirected and have neither self loops
nor multiple edges.

\subsection{The augmentation problem} Two vertices of a graph are {\em
biconnected\/} if they are in the same connected component and remain
so after the removal of any single edge or any single vertex other
than either of them.  A set of vertices is {\em biconnected\/} if
every pair of its vertices are biconnected; similarly, a graph is {\em
biconnected} if its set of vertices is biconnected.  To suit our
application of protecting sensitive information in statistical tables,
this definition for biconnectivity is slightly different from the one
used in standard textbooks. In particular, we define a connected
component of an isolated vertex to be biconnected and one with exactly
two vertices to be not biconnected.

A {\it block\/} of a graph is the induced subgraph of a maximal subset
of vertices that is biconnected.  A graph is {\it componentwise
biconnected\/} if every connected component is a block.  Throughout
this paper, $G = (A,B,E)$ denotes a bipartite graph.  A {\it legal
edge\/} of $G$ is an edge in $A \times B$ but not in $E$.  A {\it
biconnector\/} of $G$ is a set $L$ of legal edges such that $(A,B,E
\cup L)$ is componentwise biconnected.  An {\it optimal\/} biconnector
is one with the smallest number of edges.  Note that if $A =
\emptyset$ or $B = \emptyset$, $G$ is componentwise
biconnected. If $|A|=1$ and $B\not=\emptyset$ (or $|B|=1$ and
$A\not=\emptyset$), $G$ has no biconnector.  If $|A|
\geq 2$ and $|B|\geq 2$, $G$ has a biconnector.  In light of these
observations, the {\it optimal biconnector problem\/} is the
following: given $G = (A,B,E)$ with $|A| \geq 2$ and $|B| \geq 2$,
find an optimal biconnector of $G$.

The remainder of this paper assumes $|A| \geq 2$ and $|B| \geq 2$.
Also, let $n$ and $m$ be the numbers of vertices and edges in $G$,
respectively.

Given an edge subset $E'$ and a vertex subset $V'$ of $G$, $G-V'$
denotes $G$ without the vertices in $V'$ and their adjacent edges.
$G-E'$ denotes $(A,B,E-E')$, i.e., the resulting $G$ after the edges
in $E'$ are deleted.  $G \cup E'$ denotes $(A,B,E \cup E')$, i.e., the
resulting $G$ after the edges in $E'$ are added to $G$.

\subsection{Basic definitions}\label{sec:bas}
A {\it cut vertex} or {\it edge} of a graph is one whose removal
increases the number of connected components.  A {\it singular\/}
connected component is one formed by an isolated vertex.  A {\it
singular\/} block is one with exactly one vertex.  An {\em isolated}
block is one that is also a connected component.  A {\em pendant\/}
block is a singular block consisting of a vertex of degree $1$ or a
nonsingular block containing exactly one cut vertex.  Let $\Lambda(G)$
denote the set of pendant blocks of $G$.

A vertex of $G$ is {\it type $A$\/} or {\it $B$\/} if it is in $A$ or
$B$, respectively.  A block of $G$ is {\it type $A$\/} or {\it $B$\/}
if all of its noncut vertices are in $A$ or $B$, respectively; a block
is {\it type $AB$\/} if it has at least one noncut vertex in $A$ and
one in $B$.  A {\it legal pair} of $G$ is formed by two distinct
elements in $\Lambda(G)$ paired according to the following rules.
Type $A$ may pair with type $B$ or $AB$.  Type $B$ may pair with type
$A$ or $AB$.  Type $AB$ may pair with all three types.  A {\it
binding\/} edge for a legal pair is a legal edge between two noncut
vertices, one from each of the two blocks of the pair.

\begin{lemma}\
\label{lem_demanding}

\begin{enumerate}
\item 
\label{lem_demanding_2} 
A noncut vertex is in exactly one block.  Each pendant block contains
a noncut vertex.
\item 
\label{lem_block_type_2} 
A singular pendant block of $G$ is either type $A$ or $B$ while a
nonsingular pendant block is type $AB$ and has at least two vertices
from $A$ and at least two from $B$.
\item
\label{lem_binding}
There exists a binding edge for each legal pair of $G$.
\end{enumerate}
\end{lemma}
\begin{proof}
Straightforward.
\end{proof}

Let $\Lambda'\subseteq\Lambda(G)$.  A {\it legal matching\/} of
$\Lambda'$ is a set of legal pairs between elements in $\Lambda'$ such
that each element in $\Lambda'$ is in at most one legal pair.  A {\it
maximum} legal matching of $\Lambda'$ is one with the largest
cardinality possible.  $\MM{\Lambda'}$ denotes the cardinality of a
maximum legal matching of $\Lambda'$.  For a maximum legal matching of
$\Lambda'$, let
\[\RR{\Lambda'} = |\Lambda'| - 2 \MM{\Lambda'},\]
i.e., the number of elements in $\Lambda'$ that are not in the given
maximum legal matching.  Note that $\RR{\Lambda'}$ is the same for any
maximum legal matching of $\Lambda'$.

\begin{lemma}\label{lem_mat}
\begin{enumerate}
\item\label{lemma_matchings}
Let $W_1$ and $W_2$ be two disjoint nonempty sets of pendant blocks with
$\MM{W_1 \cup W_2} > 0$.  Then some $w_1 \in W_1$ and $w_2 \in W_2$
form a legal pair with $\MM{W_1 \cup W_2 -
\{w_1, w_2\}} = \MM{W_1 \cup W_2} - 1$.
\item\label{lem_match_num}
Let $n_A, n_B$ and $n_{AB}$ be the numbers of type $A$, $B$ and $AB$
pendant blocks in $\Lambda(G)$, respectively.  Then, $\RR{\Lambda(G)}
= n_A + n_B + n_{AB} - 2 \MM{\Lambda(G)}$, and
$\MM{\Lambda(G)} = \alpha + \beta + \gamma$, where $\alpha =
\min\{n_A,n_B\}$, $\beta = \min\{|n_A-n_B|,n_{AB}\}$ and $\gamma =
\lfloor\frac{n_{AB}-\beta}{2}\rfloor$.
\end{enumerate}
\end{lemma}
\begin{proof}
The first statement follows from the fact that $W_1 \cup W_2$ has a
maximum legal matching that contains a legal pair between $W_1$ and
$W_2$.  The second statement follows from the fact that a maximum
legal matching can be obtained by iteratively applying any applicable
rule below:
\label{sec:match}
\begin{itemize}
\item
If there are one unpaired type $A$ pendant block
and one unpaired type $B$ pendant block,
then we pair a type $A$ pendant block and a type $B$ one.
\item
If there is no unpaired type $B$ $($respectively, $A$$)$ pendant block
and there are one unpaired type $A$
$($respectively, $B$$)$ pendant block
and one unpaired type $AB$ pendant block, then we pair a type $A$
$($respectively, $B$$)$ pendant block with a type $AB$ one.
\item 
If all unpaired pendant blocks are type $AB$, then we pair two such
blocks.
\end{itemize}
\end{proof}

For all vertices $u \in G$,
$\DD{G}{u}$ denotes the number of
connected components in $X - \{u\}$ where $X$ is the connected
component of $G$ containing $u$.
$\CC(G)$ denotes the number of
connected components in $G$ that are not blocks.  $\BB{G}$ denotes the
number of edges in an optimal biconnector of $G$.  When $G$ is
connected, our target size for an optimal biconnector is:
\[\bound(G) = \thenumber{G}.\]

\subsection{Main results}
We first prove a lower bound on the size of an optimal biconnector and
then discuss two main results of this paper.

\begin{lemma}\label{lem_lower}\

\begin{enumerate}
\item \label{lem_alpha_biconnected}
$G$ is componentwise biconnected if and only if $\bound(G)=0$.
\item \label{theorem_lower_bound} 
$\BB{G} \geq \bound(G)$.
\end{enumerate}
\end{lemma}
\begin{proof}
Statement~\ref{lem_alpha_biconnected} is straightforward.  To prove
Statement~\ref{theorem_lower_bound}, it suffices to show $\BB{G} \geq
\MM{\Lambda(G)}+\RR{\Lambda(G)}$ and $\BB{G} \geq \max_{u \in
G}\DD{G}{u}+\CC(G)- 2$.  Let $L$ be an optimal biconnector of $G$.

To prove $\BB{G} \geq \MM{\Lambda(G)}+\RR{\Lambda(G)}$, note that
$\Lambda(G \cup L)$ is empty. Thus, every block in $\Lambda(G)$
contains an endpoint of an edge in $L$.  Since all the edges in $L$
are legal, $L$ contains at least $\MM{\Lambda(G)}+\RR{\Lambda(G)}$
edges.

To prove $\BB{G} \geq \max_{u \in G}\DD{G}{u}+\CC(G)- 2$, we need such
an $L$ that the non-block connected components of $G$ are all
contained in the same connected component of $G \cup L$.  If a given
$L$ has not yet satisfied this property, then let $X_1$ and $X_2$ be
two non-block connected components of $G$ that are contained in two
different connected components $X'_1$ and $X'_2$ of $G \cup L$,
respectively.  Let $e_1 = (u_1, v_1) \in X'_1$ and $e_2 = (u_2,v_2)
\in X'_2$ be two edges in $L$.  Such $e_1$ and $e_2$ exist because
$X_1$ and $X_2$ are not biconnected in $G$, but $X'_1$ and $X'_2$ are
biconnected in $G \cup L$. Next, let $e'_1 = (u_1, v_2)$ and $e'_2 =
(u_2, v_1)$. Then, $L' = (L - \{e_1,e_2\})\cup\{e'_1,e'_2\}$ remains
an optimal biconnector of $G$.  Also, $L'$ connects $X'_1 - \{e_1\}$
and $X'_2 - \{e_2\}$, which include $X_1$ and $X_2$.  By repeating
this endpoint switching process, we can construct a desired $L$.  With
such an $L$, we proceed to prove $\BB{G} \geq \max_{u \in
G}\DD{G}{u}+\CC(G)- 2$.  Since this claim trivially holds if $G$ is
componentwise biconnected, we focus on the case where $G$ is not
componentwise biconnected.  Then, $\DD{G}{u}$ is maximized by some $u$
that is in a non-block connected component $G$.  Let $H_u$ be the
connected component of $G \cup L$ containing $u$.  Since $G \cup L$ is
componentwise biconnected, $H_u - \{u\}$ is connected.  Then, because
$H_u - \{u\} - L$ has $\DD{G}{u}+\CC(G)- 1$ connected components, $|L|
\geq \DD{G}{u}+\CC(G)- 2$, proving our claim.
\end{proof}

The next theorem is a main result of this paper.
\begin{theorem}\label{thm_main_all_special}
If $G$ is connected, then $\BB{G} = \bound(G)$.
\end{theorem}
\begin{proof}
By Lemma~\ref{lem_lower}, $\BB{G} \geq \bound(G)$.  For ease of
understanding, the proof for $\BB{G} \leq \bound(G)$ is delayed to
Theorem~\ref{thm_main_special} in \S\ref{sec_special_case}.
\end{proof}

The next theorem generalizes Theorem~\ref{thm_main_all_special} to $G$
that may or may not be connected.
\begin{itemize}
\item 
Let $\CC_1(G)$ be the
number of connected components of $G$ that are neither isolated edges
nor blocks.  
\item 
Let $\CC_2(G)$ be the number of isolated edges; note that
$\CC(G) = \CC_1(G) + \CC_2(G)$.  
\item Let $\CC_3(G)$ be the number of
connected components that are nonsingular blocks.
\end{itemize}

\begin{theorem}\
\label{thm_main}

Case M1: $\CC_1(G) = 1$ and $\CC_2(G) =0$. Then $\BB{G} =
\bound(G)$.

Case M2:  $\CC_1(G)+\CC_2(G) \geq 2$ and $\MM{\Lambda(G)}=0$.
Then $\BB{G} = \bound(G)$.

Case M3: $\CC_1(G)+\CC_2(G) \geq 2$ and $\MM{\Lambda(G)} > 0$.
Then $\BB{G} = \bound(G)$.

Case M4: $\CC_1(G)=0$, $\CC_2(G) = 1$, and $\CC_3(G)=0$. Then
$\BB{G} = 3$.

Case M5: $\CC_1(G)=0$, $\CC_2(G)=1$, and $\CC_3(G) > 0$.  Then
$\BB{G} = 2$.

Case M6: $\CC(G)=0$. Then $\BB{G} = 0$.
\end{theorem}

\begin{proof} 

Case M1. Let $G_1$ be the connected component of $G$ that is neither an
isolated edge nor a block.  Theorem~\ref{thm_main_all_special} applies
to the case where $G_1$ contains at least two vertices in $A$ and at
least two in $B$.  Thus, we may assume without loss of generality that
$G_1$ contains exactly one vertex $u \in A$ and $r$ vertices $v_1,
v_2,\ldots,v_r \in B$ with $r \geq 2$. Note that $\bound(G)=r$.
Because $|A| > 1$ and $\CC_2(G)=0$, there is an isolated vertex $w \in
A$ or there is a nonsingular block in $G$ containing two vertices
$w_1, w_2 \in A$.  In the former case, $\{(w,v_1),\ldots,(w, v_r)\}$
is an optimal biconnector; in the latter case, $\{(w_1, v_1)\} \cup
\{(w_2,v_2),(w_2,v_3),\ldots,(w_2,v_r)\}$ is an optimal biconnector.

Case M2. Since $\MM{\Lambda(G)}=0$, we may assume without loss of
generality that all the pendant blocks are type $A$.  Note that
$\CC_2(G) = 0$, $\CC_1(G) \geq 2$ and $\bound(G) = |\Lambda(G)|$.  Let
$G_0, \ldots, G_{\CC_1(G)-1}$ be the connected components of $G$ that
are neither isolated edges nor blocks.  Since each $G_i$ has more than
two vertices, $G_i$ has a vertex $y_i \in B$. Let
$W_{i,1},\ldots,W_{i,r_i}$ be the pendant blocks of $G_i$.  Each
$W_{i,j}$ contains a noncut vertex $x_{i,j} \in A$.  The set
$\{(x_{i,j},y_{i+1\ \bmod\ \CC_1(G)}) \mid 0 \leq i < \CC_1(G) \mbox{\
and\ } 1\leq j \leq r_i\}$ is a biconnector. By
Lemma~\ref{lem_lower}(\ref{theorem_lower_bound}), this biconnector is
optimal.

Case M3.  By Lemma~\ref{lem_lower}, $\BB{G} \geq \bound(G)$.  To prove
the upper bound, let $e$ be a legal edge of $G$.  Let $G' = G \cup
\{e\}$.  We first show how to choose $e$ so that
$\bound(G')=\bound(G)-1$.  Since $\MM{\Lambda(G)} > 0$, by
Lemma~\ref{lem_mat}(\ref{lemma_matchings}), we can find a legal pair
$w_1$ and $w_2$ in different connected components with
$\MM{\Lambda(G)-\{w_1,w_2\}}=\MM{\Lambda(G)}-1$.  By
Lemma~\ref{lem_demanding}(\ref{lem_binding}), let $e$ be a binding
edge for $w_1$ and $w_2$.  Note that $\CC(G') = \CC(G) - 1$,
$\Lambda(G') = \Lambda(G)-\{w_1,w_2\}$, $\MM{\Lambda(G')} =
\MM{\Lambda(G)} - 1$, $\RR{\Lambda(G')}=\RR{\Lambda(G)}$ and $\max_{u
\in G'} \DD{G'}{u}=\max_{u \in G} \DD{G}{u}$.  Thus, $\bound(G') =
\bound(G)-1$.

This process reduces $\CC(G)$ and $\MM{\Lambda(G)}$ by $1$ each.  We
iterate this process until either (1) $\CC_1(G)+\CC_2(G) = 1$ or (2)
$\CC_1(G)+\CC_2(G) \geq 2$ and $\MM{\Lambda(G)} = 0$.  In the latter
case, we use Case M2 to complete the proof.
In the former case,
note that  we add an edge to combine two non-singular
non-biconnected connected components
into a connected component.
This new connected component is
neither an isolated edge nor a block.
Thus, $\CC_1(G) > 0$; i.e., 
$\CC_1(G) = 1$ and $\CC_2(G) = 0$ in the resulting $G$.
We then use Case M1 to complete the proof of this case.

Case M4.  Let $(r, c)$ be the isolated edge.  Let $r' \in A$ and $c'
\in B$ be two isolated vertices.  Then, $\{(r, c'), (r', c), (r',
c')\}$ is an optimal biconnector of $G$.

Case M5.  Let $G'$ be a connected component that is a nonsingular block
in $G$.  $G'$ has a vertex $r \in A$ and a vertex $c \in B$.  Let
$(r', c')$ be the isolated edge of $G$.  Then, $\{(r, c'), (r', c)\}$
is an optimal biconnector of $G$.

Case M6. This case is straightforward.
\end{proof}

\section{A matching upper bound for a connected $\mathbf{G}$}
\label{sec_special_case}
This section assumes that $G$ is connected. 

\begin{figure}[t]
\centerline{\ \psfig{figure=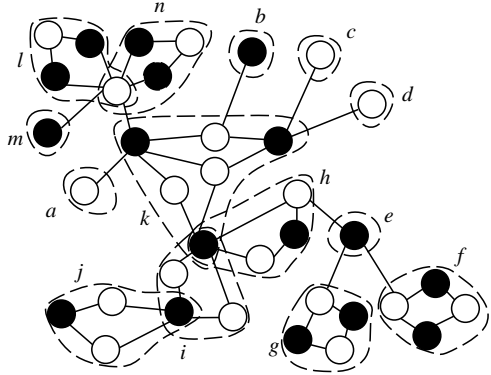,height=2in}}
\caption{
In this bipartite graph $G=(A,B,E)$, $A$ is the set of shaded vertices
and $B$ the set of unshaded vertices.  The vertices in each block of
$G$ are grouped into a dashed circle.  }
\label{fig:v2.1}
\end{figure}

\begin{figure}[t]
\centerline{\ \psfig{figure=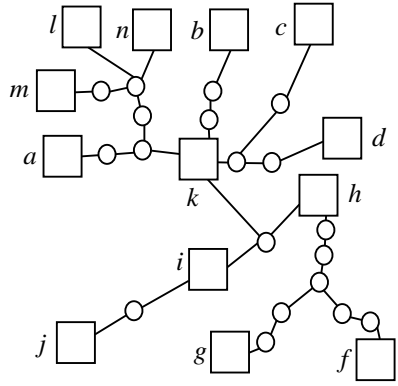,height=2in}}
\caption{
This is the block tree of the graph in
Figure~\protect{\ref{fig:v2.1}}.  The boxes are the latter's
nonsingular blocks and singular pendant ones; the circles are its cut
edges and cut vertices.}
\label{fig:v2.2}
\end{figure}

The {\em block tree\/} of $G$ is a tree $\thetree{G}$ defined as
follows.  $D_1$ denotes the set of nonsingular blocks of $G$.  $D_2$
is that of singular pendant ones.  $D_3$ is that of singular
non-pendant ones.  $C$ is that of cut vertices.  $K$ is that of cut
edges.  The vertex set of $\thetree{G}$ is $D_1 \cup D_2 \cup C \cup
K$, where $D_3$ is excluded because if $\{u\} \in D_3$, then $u \in
C$.  The vertices in $\thetree{G}$ corresponding to $D_1 \cup D_2$
are called the {\em b-vertices}; those corresponding to $C \cup K$ are
the {\em c-vertices}.  To distinguish between an edge in $G$ and one
in $\thetree{G}$, let $\bedge{y_1}{y_2}$ instead of $(y_1,y_2)$
denote an edge between two vertices $y_1$ and $y_2$ in
$\thetree{G}$.  The edge set of $\thetree{G}$ is the union of the
following sets:
\begin{itemize}
\item
$\{\bedge{d_1}{c} \mid d_1 \in D_1 \mbox{\ and\ } c \in C \mbox{\ such
that\ } c \in D_1\}$;
\item
$\{\bedge{c}{e} \mid c \in {C} \mbox{\ and\ } e \in {K}$ such that $c$
is an endpoint of $e\}$;
\item
$\{\bedge{e}{d_2} \mid e \in K \mbox{\ and\ } d_2 \in D_2
\mbox{\ such that an endpoint of\ } e \mbox{ is in } d_2\}$.
\end{itemize}
Figure~\ref{fig:v2.1} illustrates $G$ and its blocks while
Figure~\ref{fig:v2.2} illustrates its block tree.  


\begin{lemma}
\label{fact_bc_forest}
\begin{enumerate}
\item\label{fact_bc_forest_leaf_block} $\thetree{G}$ is a tree with
$O(n)$ vertices.  Its leaves are the $|\Lambda(G)|$ pendant blocks of
$G$.
\item\label{fact_bc_forest_degree} For all cut vertices $u$ in
$G$, $\DD{G}{u}$ equals the degree of $u$ in $\thetree{G}$.
\end{enumerate}
\end{lemma}
\begin{proof}
The proof is straightforward and similar to that for similar
constructs \cite{hf:graph}.
\end{proof}

Let $P_{u,v}$ denote the tree path between two vertices $u$ and $v$ in
$\thetree{G}$.  Let $|P_{u,v}|$ be the number of vertices in
$P_{u,v}$.

\begin{lemma}\label{fact_connecting_bc_noncut}
Let $Y_1$ and $Y_2$ be a legal pair of $G$.  Let $e$ be a binding edge
for $Y_1$ and $Y_2$.  Let $G'=G\cup\{e\}$.
\begin{enumerate}
\item\label{bc_1} The cut vertices of $G$ corresponding to c-vertices
in $P_{Y_1,Y_2}$ and the vertices of $G$ in the b-vertices on
$P_{Y_1,Y_2}$ form a new block $Y_e$ in $G'$.  The b-vertices of
$\thetree{G'}$ are $Y_e$ and those of $\thetree{G}$ not on
$P_{Y_1,Y_2}$.
\item\label{bc_2} The c-vertices in $\thetree{G'}$ are those in
$\thetree{G}$ excluding the ones on $P_{Y_1,Y_2}$ that are of degree $2$
in $\thetree{G}$.
\item\label{bc_3}
The edge set of $\thetree{G'}$ is the union of 
\begin{itemize}
\item the set of edges in $\thetree{G}$ whose two endpoints are
still in $\thetree{G'}$;
\item $\{\bedge{u}{Y_e} \mid u \in P_{Y_1,Y_2}$ is a cut vertex of $G$
that remains in $\thetree{G'}\}$;
\item $\{\bedge{u}{Y_e} \mid u \not\in P_{Y_1,Y_2}$ is a cut vertex of $G$
incident to $P_{Y_1,Y_2}$ in $\thetree{G}\}$.
\end{itemize}
\item\label{bc_size} The number of vertices in $\thetree{G'}$ is at
most that for $\thetree{G}$ minus $\frac{|P_{Y_1,Y_2}|-1}{2}$.
\item\label{fact_leaf_con} If $P_{Y_1,Y_2}$ contains a b-vertex of degree at
least four in $\thetree{G}$ or two vertices each of degree at least
three, then $\Lambda(G') = \Lambda(G) - \{Y_1, Y_2\}$.
\end{enumerate}
\end{lemma}
\begin{proof}
The proof is straightforward and similar to those for similar
constructs \cite{ET76b,hf:graph,hts:biconnect,RG77}.
\end{proof}

A cut vertex $u$ of $G$ is {\em massive} if $\DD{G}{u}-1>
\MM{\Lambda(G)} + \RR{\Lambda(G)}$; it is {\em critical} if
$\DD{G}{u}-1 = \MM{\Lambda(G)} + \RR{\Lambda(G)}$.

\begin{lemma}\label{lemma_structure}
Assume $\Lambda(G) > 3$.
\begin{enumerate}
\item \label{structure_2c}
$G$ has at most two critical vertices. If it has two, then
$\RR{\Lambda(G)} = 0$.
\item \label{structure_1m}
$G$ has at most one massive vertex. If it has one, then it
has no critical vertex.
\end{enumerate}
\end{lemma}
\begin{proof}
The proof follows from Lemma~\ref{fact_bc_forest}, the inequality
$\MM{\Lambda(G)} + \RR{\Lambda(G)} \geq \frac{|\Lambda(G)|}{2}$, and
basic counting arguments for trees.
\end{proof}

The next theorem is the main result of this section.
\begin{theorem}\label{thm_main_special}
$\BB{G} \leq \bound(G)$.
\end{theorem}
\begin{proof}
By Lemma~\ref{lemma_structure}, we divide the proof into the following
five cases. The first case is discussed in Lemma~\ref{lem_case0}; the
other cases are proved
in~\S\ref{subsec_case_A}--\S\ref{subsec_case_B}, respectively.

{\it Case \romanzero}: $|\Lambda(G)| \leq 3$.

{\it Case \romanone}: $|\Lambda(G)| > 3$ and $\MM{\Lambda(G)}=0$.

{\it Case \romantwo}: $|\Lambda(G)| > 3$, $\MM{\Lambda(G)}>0$, and $G$
has two critical vertices.

{\it Case \romanthree}: $|\Lambda(G)| > 3$, $\MM{\Lambda(G)}>0$, and
$G$ has no massive vertex and at most one critical vertex.

{\it Case \romanfour}: $|\Lambda(G)| > 3$, $\MM{\Lambda(G)}>0$, and
$G$ has exactly one massive vertex.
\end{proof}

\begin{lemma} \label{lem_case0}
For Case~{\romanzero}, Theorem~\ref{thm_main_special}
holds. Furthermore, given $G$, an optimal biconnector can be computed
in $O(m+n)$ time.
\end{lemma}
\begin{proof}
Straightforward.
\end{proof}

\subsection{Case {\romanone} of Theorem~\protect{\ref{thm_main_special}}}\
\label{subsec_case_A} 

\begin{lemma} \label{lem_case1}
Theorem~\ref{thm_main_special} holds for Case~{\romanone}.
\end{lemma}
\begin{proof}
Let $k = |\Lambda(G)|$. Since $\MM{\Lambda(G)} = 0$, $\bound(G) = k$
by Lemma~\ref{fact_bc_forest}. It suffices to construct a biconnector
$L$ of $k$ edges for $G$.  Let $Y_1,\ldots,Y_k$ be the pendant blocks
of $G$.  Since $\MM{\Lambda(G)}=0$, $Y_i = \{y_i\}$ and we may assume
$y_i \in B$ without loss of generality.  Then, $G$ has a cut edge
$(x_i, y_i)$ for each $y_i$, where $x_i \in A$.  Since $|A| \geq 2$
and $\MM{\Lambda(G)} = 0$, there is some $x_j \neq x_1$.  Let $G'$ be
the connected component of $G-\{x_1\}$ containing $x_j$.  Let $L$ be the
set of legal edges $(y_i, x_1)$ for all $y_i \in G'$ and $(y_i,x_j)$
for all $y_i \not\in G'$.  It is straightforward to prove that $L$ is
as desired by means of Lemma~\ref{fact_connecting_bc_noncut}.
\end{proof}

\subsection{Case {\romantwo} of Theorem
\protect{\ref{thm_main_special}}}
\label{subsec_case_D}
A path $v_1,\ldots,v_k$ in $\thetree{G}$ is {\em branchless} if for
all $i$ with $1 < i < k$ the degree of $v_i$ in $\thetree{G}$ is
two.  Let $u_1$ and $u_2$ be the critical vertices of $G$.  A leaf
{\em clings} to $u_i$ in $\thetree{G}$ if there is a branchless path
between it and $u_i$.

\begin{lemma}\
\label{lem_2_critical_match}
\begin{enumerate}
\item \label{lem_2_critical_match_2}
$\bound(G) = \MM{\Lambda(G)} = \frac{|\Lambda(G)|}{2}$.
\item \label{lem_2_critical_match_3} $\thetree{G}$ has a branchless
path between $u_1$ and $u_2$, and exactly $\frac{|\Lambda(G)|}{2}$ leaves
cling to $u_1$ only while the other $\frac{|\Lambda(G)|}{2}$ leaves cling to
$u_2$ only.
\item $\Lambda(G)$ has a maximum legal matching in which each legal
pair is between one clinging to $u_1$ and one clinging to $u_2$.
\end{enumerate}
\end{lemma}
\begin{proof}
Statement 1 follows Lemma~\ref{lemma_structure}.  Statement 2 follows
from basic counting arguments for trees.  Statement 3 follows from the
first two and Lemma~\ref{lem_mat}(\ref{lemma_matchings}).
\end{proof}

\begin{lemma} \label{lem_case2}
Theorem~\ref{thm_main_special} holds for Case~{\romantwo}.
\end{lemma}
\begin{proof}
We add to $G$ a binding edge for each legal pair in the maximum legal
matching of Lemma~\ref{lem_2_critical_match}. By
Lemmas~\ref{fact_connecting_bc_noncut} and
\ref{lem_2_critical_match}, the resulting graph is biconnected.
We add $\frac{|\Lambda(G)|}{2}$ edges, which by
Lemma~\ref{lem_2_critical_match}(\ref{lem_2_critical_match_2}) is
optimal.
\end{proof}

\subsection{Case {\romanthree} of
Theorem~\protect{\ref{thm_main_special}}}\label{subsec_case_C} Since
$|\Lambda(G)| > 3$, we can divide Case {\romanthree} into two
subcases:

{\it Case {\romanthree}-1}: $\thetree{G}$ has exactly one vertex of
degree at least three.

{\it Case {\romanthree}-2}: $\thetree{G}$ has more than one vertex of
degree at least three.

\begin{lemma}\label{lem_case3_1}
Theorem~\ref{thm_main_special} holds for Case {\romanthree}-1.
\end{lemma}
\begin{proof}
Let $x$ be the vertex in $\thetree{G}$ of degree at least three.
There are two cases:

{\case} 1: $x$ is a $b$-vertex.  Then,
$\bound(G)=\RR{\Lambda(G)}+\MM{\Lambda(G)}$.

{\case} 2: $x$ is a $c$-vertex.  Since $x$ is not massive,
$\bound(G)=\DD{G}{x}-1=\RR{\Lambda(G)}+\MM{\Lambda(G)}$ and
$\MM{\Lambda(G)}=1$.

In either case, let $N_1$ be a maximum legal matching of $\Lambda(G)$;
next, let $N_2$ be a set of legal pairs formed by pairing each pendant
block not yet matched in $N_1$ with one already matched.  Then, $N =
N_1 \cup N_2$ is a set of the smallest number of legal pairs of $G$
such that each element in $\Lambda(G)$ is in a pair.  We add to $G$ a
binding edge for each pair in $N$.  Since $\MM{\Lambda(G)}>0$, we add
$\bound(G)$ edges.  Since $\MM{\Lambda(G)}>0$ in Case 1 and
$\MM{\Lambda(G)}=1$ in Case 2, these edges form a desired biconnector
by Lemma~\ref{fact_connecting_bc_noncut}.
\end{proof}

To discuss Case {\romanthree}-2, we further assume that $\thetree{G}$
is rooted at a vertex with at least two neighbors; however, the degree
of a vertex in $\thetree{G}$ still refers to its number of neighbors
instead of children.

The next lemma chooses an advantageous root for $\thetree{G}$ for our
augmentation algorithm.  Given a vertex $v$ in $\thetree{G}$, a {\it
branch} of $v$, also called a {\em $v$-branch}, is the subtree of
$\thetree{G}$ rooted at a child of $v$.  A {\it chain} of $v$, also
called a {\em $v$-chain}, is a $v$-branch that contains exactly one
leaf in $\thetree{G}$.

Let $c^*$ be a c-vertex in $\thetree{G}$ of the largest possible
degree.
\begin{lemma}\label{lem_nice_pair}
In Case \romanthree-2, we can reroot $\thetree{G}$ at a vertex $h$
such that
\begin{enumerate}
\item \label{nice_1}
either $h$ is of degree two and no $h$-branch is a chain or $h$ is of
degree at least three;
\item \label{nice_3}
if $c^*$ is critical, then $h = c^*$.
\end{enumerate}
\end{lemma}
\begin{proof}
Let $r$ be the current root of $\thetree{G}$.  There are three cases.

{\case} 1: $c^*$ is not critical, and either $r$ is of degree two and
no $r$-branch is a chain or $r$ is of degree at least three.  We set
$h = r$.

{\case} 2: $c^*$ is not critical, $r$ is of degree two, and an
$r$-branch is a chain.  Note that $\thetree{G}$ has a vertex $r^*$ of
degree at least three.  We set $h = r^*$.

{\case} 3: $c^*$ is critical.  Since $|\Lambda(G)| > 3$, $c^*$ is of
degree three or more.  We set $h = c^*$.
\end{proof}

\begin{lemma}\label{lem_find_pair}
Let $h$ be the root of $\thetree{G}$.  In Case \romanthree-2, if
either $h$ is of degree two and no $h$-branch is a chain or $h$ is of
degree at least three, then $G$ has a legal pair $w_1$ and $w_2$ such
that
\begin{enumerate}
\item \label{find_1}
$P_{w_1,w_2}$ passes through
$h$ and two vertices of degree at least three;
\item \label{find_2}
$\MM{\Lambda(G)-\{w_1,w_2\}}=\MM{\Lambda(G)}-1$.
\end{enumerate}
\end{lemma}
\begin{proof}
There are two cases.

{\case} 1: The degree of $h$ is two and no $h$-branch
is a chain.  Let $T^*$ be an $h$-branch.

{\case} 2: The degree of $h$ is at least three.  Since this is
Case~\romanthree-2, some descendant of $h$ has degree at least three.
Let $T^*$ be the $h$-branch containing that descendant.

Let $W_1$ be the set of leaves in $T^*$.  Let $W_2=\Lambda(G)-W_1$.
By Lemma~\ref{lem_mat}(\ref{lemma_matchings}), there exist a legal
pair $w_1 \in W_1$ and $w_2 \in W_2$ with
$\MM{\Lambda(G)-\{w_1,w_2\}}=\MM{\Lambda(G)}-1$.  Then, $P_{w_1,w_2}$ contains
$h$ as desired.  Furthermore, in Case 1, $P_{w_1,w_2}$ contains a vertex of
degree at least three in $T^*$ and another in $\thetree{G} - T^*$; in
Case 2, $h$ itself is of degree at least three, and $P_{w_1,w_2}$ contain a
vertex of degree at least three in $T^*$.
In both cases, $P_{w_1,w_2}$ is as desired.
\end{proof}

\begin{lemma}\label{lem_basic_case3} In Case \romanthree-2, we can add
a legal edge to $G$ such that
\begin{enumerate}
\item \label{case3_2} 
the resulting graph $G'$ satisfies Case~{\romanzero}, {\romanone},
{\romantwo} or {\romanthree};
\item \label{case3_1} $\bound(G') = \bound(G) - 1$;
\item \label{case3_3}
if $G$ has a critical vertex, then that vertex remains critical in
$G'$.
\end{enumerate}
\end{lemma}
\begin{proof}
We use Lemma~\ref{lem_nice_pair} to reroot $\thetree{G}$, use
Lemma~\ref{lem_find_pair} to pick a legal pair $w_1$ and $w_2$, and
then add a binding edge for this pair to $G$.  By
Lemmas~\ref{lem_find_pair}(\ref{find_1}) and
\ref{fact_connecting_bc_noncut}(\ref{fact_leaf_con}),
$\Lambda(G')=\Lambda(G)-\{w_1,w_2\}$.  By
Lemma~\ref{lem_find_pair}(\ref{find_2}),
$\MM{\Lambda(G')}=\MM{\Lambda(G)}-1$.  Hence
$\MM{\Lambda(G')}+\RR{\Lambda(G')}=\MM{\Lambda(G)}+\RR{\Lambda(G)}-1$.
There are two cases.

{\case} 1: $G$ has no critical vertex.  Then, by
Lemma~\ref{fact_connecting_bc_noncut}, $\max_{u \in
G'}\DD{G'}{u}\leq\max_{u \in G}$
$\DD{G}{u}\leq\MM{\Lambda(G)}+\RR{\Lambda(G)}$.

{\case} 2: $G$ has a critical vertex. Then, $c^*$ is the critical
vertex and $\DD{G}{c^*} > \max_{u \neq c^*}\DD{G}{u}$.  By
Lemmas~\ref{lem_nice_pair}(\ref{nice_3}),
\ref{lem_find_pair}(\ref{find_1}), and
Lemma~\ref{fact_connecting_bc_noncut}, $\max_{u \in G'} \DD{G'}{u}
=\max_{u \in G}$ $\DD{G}{u}-1=\MM{\Lambda(G)}+\RR{\Lambda(G)}$.  Hence
$c^*$ remains to be a critical vertex.

In either case, $\max_{u \in G'}\DD{G'}{u} -1 \leq
\MM{\Lambda(G')}+\RR{\Lambda(G')}$.
Then, $\bound(G')=\bound(G)-1$.
Also, $G'$ has no massive vertex and thus satisfies Case~{\romanzero},
{\romanone}, {\romantwo} or {\romanthree}.
\end{proof}

\begin{lemma}\label{lem_case3} Theorem~\ref{thm_main_special} holds
for Case~{\romanthree}.
\end{lemma}
\begin{proof}
For Case~{\romanthree}-1, we use Lemma~\ref{lem_case3_1}.  For
Case~{\romanthree}-2, we add one edge to $G$ at a time using
Lemma~\ref{lem_basic_case3} until the resulting graph $G'$ does not
satisfy Case~{\romanthree}-2.  By
Lemma~\ref{lem_basic_case3}(\ref{case3_2}), $G'$ satisfies
Case~{\romanzero}, {\romanone}, {\romantwo} or {\romanthree}-1.  Thus,
we apply Lemma~\ref{lem_case0}, \ref{lem_case1}, \ref{lem_case2}, or
\ref{lem_case3_1} to $G'$ accordingly.  By
Lemma~\ref{lem_basic_case3}(\ref{case3_1}), the number of edges added
is $\bound(G)$.
\end{proof}

\subsection{Case {\romanfour} of
Theorem~\protect{\ref{thm_main_special}}}
\label{subsec_case_B}
Let $r$ be the massive cut vertex of $G$.  Let $\thetree{G}$ be rooted at $r$.

\begin{lemma}\label{lem_case_four}\ 
\begin{enumerate}
\item
\label{lem_three_dec}
$\bound(G)=\DD{G}{r}-1>\MM{\Lambda(G)}+\RR{\Lambda(G)}>\DD{G}{u} - 1$ for
any vertex $u \neq r$.
\item
\label{lem_B_r}
$\DD{G}{r} \geq 4$ and there are at least four $r$-chains.
\item \label{lemma_B_pair} The tree $\thetree{G}$ contains a legal
pair $Y_1$ and $Y_2$ as well as two distinct $r$-branches $T_1$ and
$T_2$ such that $T_1$ is a chain, $Y_1 \in T_1$, and $Y_2 \in T_2$.
\end{enumerate}
\end{lemma}
\begin{proof}

Statement \ref{lem_three_dec}. This statement follows from the
definition of Case {\romanfour}.

Statement \ref{lem_B_r}.  Let $\delta_1$ be the number of
$r$-chains. Then, $\DD{G}{r} \geq \delta_1$ and $|\Lambda(G)| \geq 2
(\DD{G}{r} - \delta_1) + \delta_1$.  So $\DD{G}{r} \geq \delta_1 \geq
2 \DD{G}{r} - |\Lambda(G)|$.  Let $\delta_2 = (\DD{G}{r}-1) -
(\MM{\Lambda(G)} + \RR{\Lambda(G)})$.  Because $r$ is massive,
$\delta_2 \geq 1$.  Note that $|\Lambda(G)| = 2 \MM{\Lambda(G)} +
\RR{\Lambda(G)}$.  Thus $\DD{G}{r} \geq \delta_1 \geq 2 \delta_2 + 2 +
\RR{\Lambda(G)} \geq 4$.

Statement~\ref{lemma_B_pair}. Let $T_1$ be an $r$-chain.  Let $Y_1$ be
the leaf of $\thetree{G}$ in $T_1$.  Because $\MM{\Lambda(G)}>0$,
$\thetree{G}$ contains a leaf $Y_2 \neq Y_1$ that forms a legal pair
with $Y_1$.  Let $T_2$ be the $r$-branch that contains $Y_2$.  Then,
$Y_1$, $Y_2$, $T_1$ and $T_2$ are as desired.
\end{proof}

\begin{lemma}\label{lem_basic_case4}
We can add a legal edge to $G$ such that for the resulting graph $G'$,
\begin{enumerate}
\item \label{case4_1}
$\bound(G') = \bound(G) - 1$;
\item \label{case4_2}
$\DD{G'}{r} = \DD{G}{r} - 1$.
\end{enumerate}
\end{lemma}
\begin{proof}
Let $Y_1$, $Y_2$, $T_1$ and $T_2$ be as stated in
Lemma~\ref{lem_case_four}(\ref{lemma_B_pair}).  The added edge is a
binding edge for $Y_1$ and $Y_2$.  By
Lemma~\ref{fact_connecting_bc_noncut}, the b-vertices and c-vertices on
$P_{Y_1,Y_2}$ form a new block $Y'$ in $G'$.  $Y'$ may or may not be a leaf in
$\thetree{G'}$; in either case, $\MM{\Lambda(G')}+\RR{\Lambda(G')}
\leq \MM{\Lambda(G)}+\RR{\Lambda(G)}$.
Note that $P_{Y_1,Y_2}$ contains $r$.  Thus, by
Lemmas~\ref{fact_connecting_bc_noncut} and
\ref{lem_case_four}(\ref{lem_B_r}), $r$ remains a cut vertex in
$G'$ with $\DD{G'}{r}=\DD{G}{r}-1$ while $\DD{G'}{v}\leq\DD{G}{v}$ for
all vertices $v \neq r$.  Consequently, $\bound(G')=\bound(G)-1$.
\end{proof}

\begin{lemma}\label{time_VI}
Theorem~\ref{thm_main_special} holds for Case~{\romanfour}.  Moreover,
this case can be reduced in linear time to Case {\romanzero},
{\romanone}, {\romantwo} or {\romanthree}.

\end{lemma}
\begin{proof}
We add one edge to $G$ at a time using Lemma~\ref{lem_basic_case4}
until the resulting graph $G'$ satisfies Case {\romanzero},
{\romanone}, {\romantwo} or {\romanthree}.  Thus, we apply
Lemma~\ref{lem_case0}, \ref{lem_case1}, \ref{lem_case2} or
\ref{lem_case3} accordingly.  By
Lemma~\ref{lem_basic_case4}(\ref{case4_1}), $\bound(G)$ edges are
added.  To implement this proof in linear time, we first define a data
structure as follows.

Let $Q$ be the set of leaves of $\thetree{G}$ that are in the
$r$-chains.  We set up a counter for the number of these leaves.  We
also set up three doubly linked lists containing those of them that
are types $A$, $B$, and $AB$, respectively.

We set up a counter for the number of $r$-branches that are not chains.
For each such branch, we set up a doubly linked list for the leaves of
$\thetree{G}$ in it.  We also set up three doubly linked lists for the
leaves in these branches that are types $A$, $B$, and $AB$,
respectively.

Given $G$, we can set up these linked lists and counters in linear
time.  We next use this data structure to find a legal pair $Y_1$ and
$Y_2$ by means of Lemma~\ref{lem_case_four}(\ref{lemma_B_pair}).
Since $|Q| \geq 4$ by Lemma~\ref{lem_case_four}(\ref{lem_B_r}), there
are two cases.

{\case} 1: Some $Y_1$ and $Y_2 \in Q$ form a legal pair. This is our
desired pair.  Note the $r$-chains containing $Y_1$ and $Y_2$ in
$\thetree{G}$ are contracted into a new chain in $\thetree{G'}$
consisting of a single leaf of type $AB$.  

{\case} 2: $Q$ contains only type $A$ or $B$ leaves.  Select any $Y_1
\in Q$.  Since $\MM{\Lambda(G)} > 0$, some $Y_2 \in \Lambda(G) - Q$
forms a desire legal pair with $Y_1$.  Note that $Y_1$ and $Y_2$ are
no longer pendant blocks in $G'$ and the newly created block is not a
pendant block of $G'$, either.  The $r$-branch containing $Y_2$
becomes a chain if in $G$ it contains exactly two pendant blocks.

It takes $O(1)$ time to decide which of these two cases holds.  In
either case, the selection of $Y_1$ and $Y_2$ takes in $O(1)$ time
using the linked lists.  Once $Y_1$ and $Y_2$ are found, we can find a
binding edge in $O(1)$ time in a straightforward manner.  After the
edge is added to $G$, we can update the data structure in $O(1)$ time
for $G'$.  Then, we use Lemma~\ref{lem_mat}(\ref{lem_match_num})
and the counters to
check whether $G'$ satisfies Case~{\romanfour} in $O(1)$ time.
We repeat this process until $G'$ does not satisfies
Case~{\romanfour}. At this point, we complete the reduction.
Since we iteratively add at most $O(n)$ edges in
Case~{\romanfour}, the reduction takes linear time.
\end{proof}

\section{Computing an optimal biconnector in linear time}\
\label{sec_linear_time}

\begin{theorem}\label{thm_linear_time}
Given $G$, an optimal biconnector is computable in $O(m+n)$ time.
\end{theorem}

We prove this theorem by means of Theorems~\ref{thm_main} and
\ref{thm_main_special} as follows.

Given $G$, it takes $O(m+n)$ time to determine which case of
Theorem~\ref{thm_main} holds. Then, it takes $O(m+n)$ time in a
straightforward manner to compute an optimal biconnector for Cases M2,
M4, M5 and M6; reduce Case M3 to Case M1 or M2; and reduce Case M1 to
Theorem~\ref{thm_main_special}.

Next, it takes $O(m+n)$ time to determine which case of
Theorem~\ref{thm_main_special} holds.  Then, it is straightforward to
compute an optimal biconnector in $O(m+n)$ time for Cases
{\romanzero}, {\romanone}, and {\romantwo}.  Lemma~\ref{time_VI}
reduces Case {\romanfour} in $O(m+n)$ time to Case {\romanzero},
{\romanone}, {\romantwo} or {\romanthree}.  By
Lemma~\ref{lem_case3_1}, we can find an optimal biconnector in
$O(m+n)$ time for Case {\romanthree}-1.  The remaining proof shows how
to reduce Case {\romanthree}-2 to Case {\romanzero}, {\romanone},
{\romantwo} or {\romanthree}-1 in $O(m+n)$ time by implementing the
proof of Lemma~\ref{lem_case3}.

We define a data structure $\hht{G}$ as follows.  First, we root
$\thetree{G}$ at a vertex of degree two or more as in
\S\ref{subsec_case_C} and classify each vertex $u$ by a 4-bit
code $\sigma_0\sigma_1\sigma_2\sigma_3$ based on the subtree $T_u$ of
$\thetree{G}$ rooted at $u$:
\begin{itemize}
\item $\sigma_0 = 1$  if and only if $T_u$ has more than one leaf;
\item  $\sigma_1$, $\sigma_2$ or $\sigma_3 = 1$
if and only if $T_u$ contains a leaf
of type $A$, $B$ or $AB$, respectively.
\end{itemize}
The code has at most ten combinations, i.e., $0100$, $0010$, $0001$ and
all the combinations with $\sigma_0=1$ except $1000$.
$\hht{G}$ is
$\thetree{G}$ augmented with the following items:
\begin{enumerate}
\item\label{data_l}
At each vertex in $\thetree{G}$, $\hht{G}$ maintains
its degree and
a doubly linked list for the children of $u$ with the
same $\sigma_0\sigma_1\sigma_2\sigma_3$ code.
There are ten such lists.
\item\label{data_ab} 
There are three counters for the numbers of leaves in $\thetree{G}$ of
types $A$, $B$ and $AB$, respectively.
\item\label{data_c}
The c-vertices of degree at least three are partitioned into groups of
the same degree.  Each nonempty group is arranged into a doubly linked
list.  The lists themselves are connected by a doubly linked list in
the increasing order of vertex degrees.
\end{enumerate}
We do not need parent pointers in $\hht{G}$, which are subtle to
update \cite{hts:triconnect,hts:biconnect,RG77}.  This finishes the
description of $\hht{G}$.  We can build $\hht{G}$ from $G$ in $O(m+n)$
time.

\begin{lemma}\label{lem_time}
\begin{enumerate}
\item\label{time_reroot} Let $r$ be the current root of $\hht{G}$.
Let $h$ be as stated in Lemma~\ref{lem_nice_pair}.  Given $\hht{G}$,
if $r$ is critical, we can reroot $\hht{G}$ in $O(1)$ time according to
Lemma \ref{lem_nice_pair}; $O(n)$ time if $r$ is not critical but $h$
is; or $O(|P_{r,h}|)$ time if neither is.

\item\label{time_find}
Given $\hht{G}$, we can find $w_1$ and $w_2$ of
Lemma~\ref{lem_find_pair} in $O(|P_{w_1,w_2}|)$ time.
\end{enumerate}
\end{lemma}
\begin{proof}

Statement \ref{time_reroot}.
We implement the proof Lemma~\ref{lem_nice_pair} using the following steps.
\begin{enumerate}
\item\label{step_r1}
Use Item \ref{data_c} of $\hht{G}$ to find $c^*$.
\item\label{step_r2}
Use Items \ref{data_l} and \ref{data_ab} of $\hht{G}$ and
Lemma~\ref{lem_mat}(\ref{lem_match_num}) to decide which case of the
proof of Lemma~\ref{lem_nice_pair} holds.
\item\label{step_r3}
\begin{enumerate}
\item\label{step_a0}
For Case 1 of the proof of Lemma~\ref{lem_nice_pair}, set $h=r$ and
$\hht{G}$ is unchanged.
\item\label{step_a1}
For Case 2 of the proof of Lemma~\ref{lem_nice_pair}, first use Item
\ref{data_l} of $\hht{G}$ to find the nearest desired 
descendant $r^*$ of $r$ and then reroot $\hht{G}$ at $h=r^*$ and
update it accordingly.
\item\label{step_a2}
For Case 3 of the proof of Lemma~\ref{lem_nice_pair}, if $r \neq c^*$,
then recompute $\hht{G}$ from $\thetree{G}$ to root at $h=c^*$;
otherwise, $r=c^*$, and $\hht{G}$ is unchanged.
\end{enumerate}
\end{enumerate}

Since Steps \ref{step_r1} and \ref{step_r2} take $O(1)$ time, the the
time complexity of each case of this statement is bounded by that of Step 
\ref{step_r3}. 

{\case} 1: $r$ is critical. Step~\ref{step_a2} runs with $r = c^*$ in
$O(1)$ time.

{\case} 2: $r$ is not critical but $h$ is.  Step~\ref{step_a2} runs
with $r \neq c^*$ in $O(n)$ time.

{\case} 3: Neither $r$ nor $h$ is critical.  Then, Step~\ref{step_a0}
or Step~\ref{step_a1} is performed.  Step~\ref{step_a0} takes $O(1)$
time.  For Step~\ref{step_a1}, the search for $r^*$ takes $O(1)$ time
per vertex on $P_{r,r^*}$. Since the internal vertices of $P_{r,r^*}$
all have degree two, updating Item~\ref{data_l} of $\hht{G}$ along
this path takes $O(1)$ time per vertex.  Item~\ref{data_l} of
$\hht{G}$ outside this path and the other two items remain the
same. Thus, this case takes $O(P_{r,h})$ total time as desired.

Statement \ref{time_find}.  We implement the proof of
Lemma~\ref{lem_find_pair} using the following steps.

\begin{enumerate}
\item\label{step_bbc}
Use Item~\ref{data_l} of $\hht{G}$ to decide which case of the proof of 
Lemma~\ref{lem_find_pair} holds.
\item\label{step_b}
Use Item~\ref{data_ab} of $\hht{G}$ and
Lemma~\ref{lem_mat}(\ref{lem_match_num}) to find all possible pairs of
types $t_1$ and $t_2$ such that $\Lambda(G)$ has a maximum matching
that contains a legal pair between type $t_1$ and type $t_2$.
\item\label{step_ba} 
For each such pair of $t_1$ and $t_2$, perform the following
computation until $w_1$ and $w_2$ are found.
\begin{enumerate}
\item\label{step_b2}
For Case 1 of the proof of Lemma~\ref{lem_find_pair}, $w_1$ and $w_2$
are in the two branches of the root of $\hht{G}$ separately.  Use
Item~\ref{data_l} of $\hht{G}$ at the root to decide whether the
desired $w_1$ and $w_2$ exist. If they exist, use Item~\ref{data_l} of
$\hht{G}$ to search for them.
\item\label{step_b1}
For Case 2 of the proof of Lemma~\ref{lem_find_pair}, $w_1$ and $w_2$
are in two separate branches of the root, of which one is not a chain.
The remaining computation is similar to that of Step~\ref{step_b2}.
\end{enumerate}
\end{enumerate}
By Lemma~\ref{lem_nice_pair}, some pair $t_1$ and $t_2$ yields the
desired $w_1$ and $w_2$.  Steps \ref{step_bbc} and \ref{step_b} take
$O(1)$ time. There are $O(1)$ possible pairs of $t_1$ and $t_2$.  For
each such pair, checking the existence of $w_1$ and $w_2$ takes $O(1)$
time. If they exist, searching for them takes $O(1)$ time per vertex
on the path $P_{w_1,w_2}$.
\end{proof}

The next lemma completes the proof of Theorem~\ref{thm_linear_time}.
\begin{lemma}
Case {\romanthree}-2 is reducible to Case {\romanzero}, {\romanone},
{\romantwo} or {\romanthree}-1 in $O(m+n)$ time.
\end{lemma}
\begin{proof}
Given $G$ in Case {\romanthree}-2 as input, the reduction algorithm is
as follows:
\begin{enumerate}
\item\label{step_1} Construct $\hht{G}$.
\item\label{step_2} {\bf repeat} 
\begin{enumerate}
\item	\label{step_a}
Use Lemma~\ref{lem_time}(\ref{time_reroot}) to reroot $\hht{G}$.
\item	\label{step_a22}
Use Lemma~\ref{lem_time}(\ref{time_find})
to find a legal pair $w_1$ and $w_2$.
\item\label{step_bx} Add a binding edge $e$ for $w_1$ and $w_2$ into $G$.
\item\label{step_c} 
Use Lemma~\ref{fact_connecting_bc_noncut} to update $\hht{G}$ while
rerooting it at the new b-vertex $Y_e$ resulting from the insertion of
$e$.
\end{enumerate}
{\bf until} $G$ does not satisfy Case {\romanthree}-2.
\end{enumerate}

Since Step~\ref{step_1} takes $O(m+n)$ time, it suffices to prove that
Step~\ref{step_2} takes $O(n)$ time. By
Lemma~\ref{lem_find_pair}(\ref{find_1}), each iteration of
Step~\ref{step_2} reduces $|\Lambda(G)|$ by two.  Since $|\Lambda(G)|
< n$, the repeat loop has less than $n$ iterations. Then, since the
until condition can be checked in $O(1)$ time per iteration using
Lemma~\ref{lem_mat}(\ref{lem_match_num}) and Items \ref{data_ab} and
\ref{data_c} of $\hht{G}$, the until step takes $O(n)$ total time.
Similarly, Step~\ref{step_bx} takes $O(1)$ time per iteration and
$O(n)$ total time in a straightforward manner.

We next show that Steps~\ref{step_a}, \ref{step_a22} and \ref{step_c}
also take $O(n)$ total time.  For a given iteration, let $G_0$ and
$G_1$ denote $G$ before and after $e$ is inserted, respectively.

Step~\ref{step_a}.  We show that each case in the proof of
Lemma~\ref{lem_time}(\ref{time_reroot}) takes $O(n)$ total time as
follows.

Case 1.  This case takes $O(1)$ time per iteration and thus $O(n)$
total time.

Case 2.  By Lemma~\ref{lem_basic_case3}(\ref{case3_3}), this case can
only happen once in the above augmentation algorithm.  Hence, this
case takes $O(n)$ total time.

Case 3.  This case takes $O(1)$ time per edge on $P_{r,h}$ for an
iteration.  Note that the degree of a vertex in $\hht{G}$ never
increases by edge insertion.  Then, since $\hht{G_1}$ is rooted at
$Y_e$ with $e$ connecting two leaves of $\hht{G_0}$, each edge on
$P_{r,h}$ is traversed only once to reroot $\hht{G}$ for this case
throughout all the iterations. Therefore, this case takes $O(n)$ total
time.

Step \ref{step_a22}. This step takes $O(|P_{w_1,w_2}|)$ time per
iteration.  Since there are $O(n)$ iterations, by
Lemma~\ref{fact_connecting_bc_noncut}(\ref{bc_size}), this step takes
$O(n)$ total time.

Step \ref{step_c}.  We bound the time for updating each item of
$\hht{G}$ as follows.

Item~\ref{data_l} of $\hht{G}$.  Notice that $P_{w_1,w_2}$ passes
through the root of $\hht{G_0}$. Also, $\hht{G_1}$ is rooted at $Y_2$.
These properties make it straightforward to update this item in
$O(|P_{w_1,w_2}|)$ time per iteration.  Since there are $O(n)$
iterations, by Lemma~\ref{fact_connecting_bc_noncut}(\ref{bc_size}),
this step takes $O(n)$ total time.

Item~\ref{data_ab} of $\hht{G}$.  By
Lemma~\ref{lem_find_pair}(\ref{find_1}),
$\Lambda(G_1)=\Lambda(G_0)-\{w_1,w_2\}$.  Thus it takes $O(1)$ time to
update this item per iteration and $O(n)$ total time.

Item~\ref{data_c} of $\hht{G}$.  Let $u$ be a c-vertex in $\hht{G_0}$.
If $u \not\in P_{w_1,w_2}$, it has the same degree in $\hht{G_0}$ and
$\hht{G_1}$ and is not relocated in this item.  If $u \in
P_{w_1,w_2}$, its degree reduces at most 2 in $\hht{G_1}$ and can be
relocated in $O(1)$ time.  Therefore, this item can be updated in
$O(|P_{w_1,w_2}|)$ time per iteration, i.e., $O(n)$ total time as
shown for Item~\ref{data_l}.
\end{proof}

\section*{Acknowledgments}
We are very grateful to Dan Gusfield for insightful discussions and to
the anonymous referee for extremely thorough comments.

%\bibliography{all}
%\bibliographystyle{siam}

\end{document}